\documentclass{aa}
\usepackage[varg]{txfonts}

\usepackage{natbib}
\bibpunct{(}{)}{;}{a}{}{,}

\begin{document}

\title{Extreme ionised outflows are more common when the radio emission is compact in AGN host galaxies} 

\subtitle{}

\author{S. J. Molyneux \inst{1}
  \thanks{E-mail: smolyneux.astro@gmail.com}\fnmsep 
\and C. M. Harrison \inst{1}
\thanks{E-mail: c.m.harrison@mail.com}\fnmsep
\and M. E. Jarvis \inst{2,1,3}}

\institute{European Southern Observatory, Karl-Schwarzschild-str. 2, 85748 Garching, Germany
\and Max-Planck Institut f{\"u}r Astrophysik, Karl-Scwarzschild-Str. 1, 85748 Garching, Germany
\and Ludwig Maximilian Universit{\"a}t, Professor-Huber-Platz 2, 80539 Munich, Germany
}

\date{.}

\abstract{Using a sample of 2922 $z$$<$$0.2$, spectroscopically-identified Active Galactic Nuclei (AGN) we explore the relationship between radio size and the prevalence of extreme ionised outflows, as traced using broad [O~{\sc iii}] emission-line profiles in Sloan Digital Sky Survey (SDSS) spectra. To classify  radio sources as compact or extended, we combine a machine-learning technique of morphological classification with size measurements from two-dimensional Gaussian models to data from all-sky radio surveys. We find that the two populations have statistically different [O~{\sc iii}] emission-line profiles, with the compact sources tending to have the most extreme gas kinematics. When the radio emission is confined within 3$^{\prime\prime}$ (i.e., within the spectroscopic fibre or $\lesssim$5\,kpc at the median redshift), there is twice the chance of observing broad [O~{\sc iii}] emission-line components, indicative of very high velocity outflows, with FWHM$>$1000\,km\,s$^{-1}$. This difference is most enhanced for the highest radio luminosity bin of $\log[L_{\rm1.4GHz}$/W\,Hz$^{-1}$]=23.5--24.5 where the AGN dominate the radio emission; specifically, $>$1000\,km\,s$^{-1}$ components are almost four-times as likely when the radio emission is compact in this subsample. Our follow-up $\approx$0.3--1\,arcsec resolution radio observations, for a subset of targets in this luminosity range, reveal that radio jets and lobes are prevalent, and suggest that compact jets might be responsible for the enhanced outflows in the wider sample. Our results are limited by the available, relatively shallow, all-sky radio surveys; however, forthcoming surveys will provide a more complete picture on the connection between radio emission and outflows. Overall, our results add to the growing body of evidence that there is a close connection between ionised outflows and compact radio emission in highly accreting `radiative' AGN, possibly due to young or frustrated, lower-power radio jets.}  

\maketitle

\section{Introduction}

Understanding the physical processes that connect galaxy growth and the growth of their central supermassive black holes remains one of the biggest outstanding problems of galaxy evolution research \citep[e.g.,][]{Alexander12,Kormendy13,Heckman14,King15,Harrison17}. The sites of growing supermassive black holes are observationally identified as ``Active Galactic Nuclei'' due to the enormous amounts of energy that they release across the electromagnetic spectrum. Cosmological models and simulations of galaxy evolution require some fraction of this released energy to couple to the surrounding interstellar medium (ISM) in order to reproduce the observed properties of massive galaxies and the surrounding intergalactic medium \citep[IGM; e.g.,][]{Vogelsberger14,Hirschmann14,Crain15,Beckmann17,Choi18}. However, the details of how this process works in the real Universe are not well established, particularly during periods of rapid black hole growth \citep[e.g.,][]{Harrison17}.

One observational strategy to establish the connection between supermassive black hole growth and galaxy evolution is to identify, and to characterise, AGN-driven outflows in the multi-phase ISM \citep[e.g.][]{Veilleux05,Morganti05,Carniani15,Perna15,Brusa15,Balmaverde16,Rupke17,Lansbury18,Fluetsch19,RamosAlmeida19}. Of most relevance for this study, is the presence of broad and-or asymmetrical [O~{\sc iii}]$\lambda$5007 emission-line profiles which have long been used to trace outflows of warm ($\approx$10$^{4}$\,K) ionised gas in the narrow-line region of AGN \citep[][]{Heckman84,Whittle85}. These outflows can be located on $\approx$10\,pc --10\,kpc scales \citep[][]{Harrison14,Husemann16,Rupke17,VillarMartin17,Finlez18}. This is a particularly useful outflow tracer because, through large optical spectroscopic surveys such as the Sloan Digital Sky Survey \citep[SDSS;][]{York00}, measurements on large samples of $z\lesssim$0.8 AGN can be obtained  \citep[e.g.,][]{Mullaney13,Woo16,Zakamska14,Zakamska16,Balmaverde16}. Recent and on-going near-infrared spectroscopic surveys make it possible to obtain similar constraints on large samples of $z\approx1$--3 galaxies \citep[e.g.,][]{Harrison16,Leung17,ForsterSchreiber19}.

Here we are particularly interested in the observations showing that the prevalence and/or velocities of [O~{\sc iii}] outflows is related to the radio luminosity \citep[][]{Mullaney13,VillarMartin14,Zakamska14,Hwang18}; namely, that there is a higher prevalence of the most powerful outflows when the radio luminosity is higher. Although some work does not find such a relationship, they can suffer from a relatively high radio detection limit; for example the study of \citet{Woo18} is limited to radio luminosities of $\log[L_{\rm1.4GHz}$/W\,Hz$^{-1}]\gtrsim$24. Indeed, the relationship between outflows and radio emission, may even be strongest for AGN with moderate to intermediate radio luminosities \citep[i.e., $\log L_{\rm1.4GHz}$/W\,Hz$^{-1}\approx$23--25; see discussion in e.g.,][]{Mullaney13,Zakamska16,Jarvis19}. 

For the most radiatively luminous AGN, sometimes called ``quasars'', it is often assumed that the dominating radiative power of the central source drives the observed outflows \citep[e.g.,][]{FaucherGiguere12,King15}. However, the observed relationship between outflows and radio emission, opens up the possibility that the mechanical power of radio jets may be a crucial outflow driving mechanism in these systems \citep[e.g.,][]{Mullaney13}. Indeed, from low-power AGN through to the most extreme sources, radio jets are seen to interact with the ISM \citep[e.g.,][]{Whittle86,Ferruit98,Tadhunter14,Riffel14,Kharb17,Nesvadba17,Finlez18,Morganti18,Jarvis19}. On the other hand, star-formation driven outflows and quasar winds that shock the ISM provide alternative possibilities for producing the observed radio emission and correlation with outflow properties \citep[e.g.,][]{Condon13,Nims15,Zakamska16,Hwang18,Panessa19}.

A tentative result that may shed more light on the outflow-radio connection is presented by \citet{Holt08}, who find that [O~{\sc iii}] outflows are more extreme in {\em compact} radio galaxies (radio emission confined within $\lesssim$10\,kpc), compared to {\em extended} radio galaxies. The authors propose that radio jets, early in their evolution, are strongly interacting with the ISM in the nuclear regions \citep[e.g.,][]{vanBreugel84,ODea91,Bicknell18,Mukherjee18}.
However, the primary sample of \citet{Holt08} contains only 14 sources, all of which represent the most powerful - and rare - radio AGN ($\log[L_{\rm 5GHz}$/W\,Hz$^{-1}]$)$>$26.4) and, furthermore, the comparison samples are also small and inhomogeneous. It is therefore not clear how applicable this result is to the bulk of the AGN population. Here we test if this result holds for more typical AGN which do not have extreme radio luminosities using a sample of $\approx$3000 targets with both [O~{\sc iii}] emission-line profile measurements and radio size measurements. 

In Section\,2 we outline the sample selection, catalogues and the radio data used. In Section\,3 we describe our analysis. In Section\,4 we discuss our results and in Section\,5 we give our conclusions. We adopt $H_0=70$\,km$^{-1}$\,Mpc$^{-1}$, $\Omega_M=0.3$, and $\Omega_\Lambda=0.7$.

\section{The Sample, Catalogues and Radio Data}

We aim to explore the relationship between the size of radio emission and the presence of ionised outflows in AGN host galaxies. To do this we make use of the valued-added spectroscopic catalogue of $\approx$24,000 AGN that were identified from the Sloan Digital Sky Survey (SDSS, Data Release 7; \citealt{Abazajian09}) that is presented in \citet{Mullaney13}\footnote{https://sites.google.com/site/sdssalpaka/downloads}  and is consequently cross-correlated with all-sky radio surveys. 

\subsection{Catalogues and sample selection}
\label{sec:sample}
The parent catalogue of \cite{Mullaney13} contains 24624 sources that were identified as AGN from optical spectroscopy, using a combination of emission-line flux ratios (`BPT' diagnostics; \citealt{Baldwin81}; to identify `Type 2' AGN) and the presence of broad H$\alpha$ emission-line components (to identify `Type 1' AGN). For each AGN, the emission-lines, including the [O~{\sc iii}]$\lambda$5007 line, were fit with two Gaussian components in order to search for broad emission-line components indicative of ionised outflows. The final sample of AGN was cross-matched, by \cite{Mullaney13}, to the 1.4\,GHz radio surveys of FIRST (\citealt{Becker95}) and NVSS (\citealt{Condon98}) largely following the procedure outlined in \cite{Best05} but including sources down to signal-to-noise ratios $>$3 in NVSS (i.e., 1.4\,GHz flux densities of $\approx$2\,mJy). Here, we make use of the matched catalogue as our `parent sample'; however, we update the FIRST radio measurements using the latest, and final, catalogue that is presented in \cite{Helfand15} and contains radio sources with signal-to-noise ratios of $\ge$5 (detection limit $\approx$1\,mJy). 

\begin{figure}
	\includegraphics[width=\columnwidth]{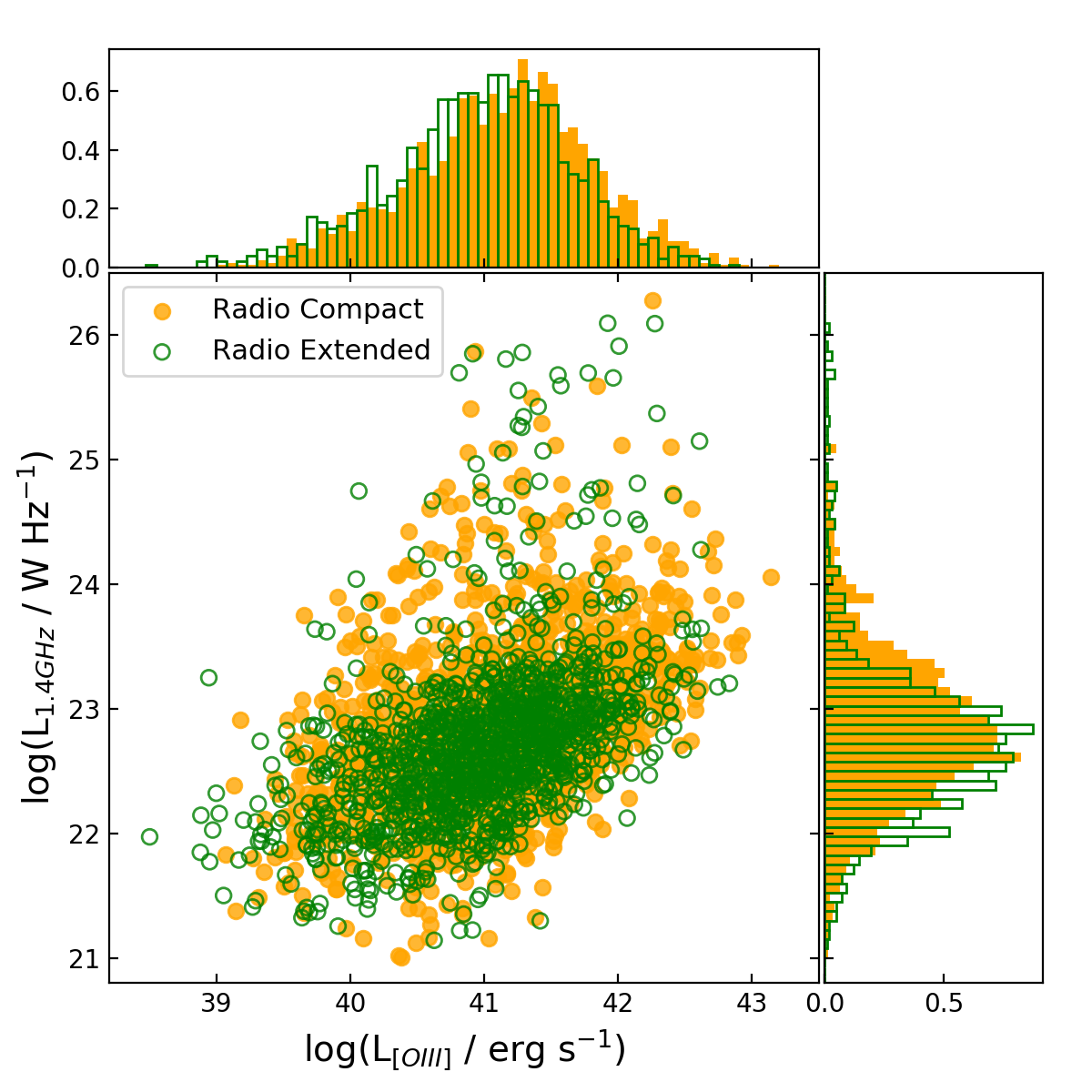}
    \caption{Radio luminosity (rest-frame 1.4\,GHz) versus [O~{\sc iii}] luminosity for our final sample of 2922 AGN. The sources classified as having extended radio emission are shown using green hollow symbols and those classified as radio compact are shown using filled orange symbols. The normalised histograms of luminosities for both populations are also shown as hollow and filled histograms, respectively. The radio and [O~{\sc iii}] luminosities of the two samples are broadly similar; nonetheless, we take into account these small differences when comparing the two populations in Section~\ref{sec:prevalenceResults}.} 
    \label{fig:RadOiiiLum}
\end{figure}

As in \cite{Mullaney13} we prefer to use the NVSS flux density measurements to infer the {\em total} radio luminosity densities ($L_{\rm 1.4GHz}$) because the larger beam (full-width-at-half-maximum [FWHM]$\approx$45\,arcsec), compared to FIRST, reduces the chances of resolving away flux or missing extended radio structures. However, the FIRST data with a resolution of $\approx$5\,arcsec is used for more accurate positional matching to the SDSS sources (\citealt{Mullaney13}). Furthermore, we also make use of superior spatial resolution of the FIRST data for radio size measurements and morphological classifications (see Section~\ref{sec:radioClassification}). 

Starting with the parent catalogue, we created the final sample to be used in this work by applying the following criteria:

\begin{enumerate}
\item We only consider AGN within a redshift range of $0.02$$<$$z$$<$$0.2$ (discussed in more detail below), leaving a sample of 16326 AGN.
\item We only consider AGN which have 1.4\,GHz radio detections in the FIRST or NVSS catalogues (required to characterise the spatial extent of the radio emission), leaving a sample of 2948 AGN. 
\item We removed a small number of sources (only 0.9\% of the sample) where, either: (a) the two-component fits to the [O~{\sc iii}] emission-line profiles failed to converge in the \cite{Mullaney13} catalogue (3 targets removed); or (b) the NVSS-only detections (i.e., those not in the FIRST catalogue) are not covered by FIRST imaging at all (which is required for our later analyses; Section~\ref{sec:radioClassification}) or that, by visual inspection, were associated with the wrong optical counterpart (usually mergers; 23 targets removed). This results in our final sample of 2922 AGN.
\end{enumerate}

The upper bound of the redshift range in step (1) is a compromise between keeping a significant number of luminous AGN in the sample, whilst having a reasonable detection limit on the radio (i.e., a limit of $\log[L_{\rm 1.4 GHz}$/W\,Hz$^{-1}]$$\approx$$22.8$ at the highest redshift of $z=0.2$) and a reasonable physical resolution (1\,arcsec corresponds to $\le$3.3\,kpc for $z\le0.2$). The lower bound of $z$=0.02 is such that the 3\,arcsec SDSS fibre still covers a reasonable fraction of the galaxy compared to the higher redshift sources (1\,arcsec corresponds to 0.4\,kpc at $z=0.02$). We further consider the varying physical size scales covered by the SDSS spectroscopy during our results presented in Section~\ref{sec:prevalenceResults}. 

The final sample used in this work consists of 2922 AGN. The distribution of rest-frame 1.4\,GHz radio luminosities and the [O~{\sc iii}] luminosities are presented in Figure~\ref{fig:RadOiiiLum}.\footnote{ We note that assuming an extreme range of spectral indices $\alpha=-0.8$ to $\alpha=1.1$ \citep[][]{Lal10} has the impact of changing the radio luminosities by a median factor of $-$16\% to $+$3\% (maximum of $-$34\% to $+$6\%). This choice does not affect our main conclusions.} We assume a radio spectral index of $\alpha=0.8$ when calculating the radio luminosity, where $F_{\nu}\propto\nu^{-\alpha}$, motivated by multi-frequency radio observations of a subset of this sample \citep[][]{Jarvis19}. Overall the sample covers five orders of magnitude in radio luminosity, with a median of $\log[L_{\rm 1.4 GHz}$/W\,Hz$^{-1}]$$=$$22.7$, and four orders of magnitude in [O~{\sc iii}] luminosity, with a median of $\log[L_{\rm [O III]}$/erg\,s$^{-1}]$$=$$41.1$. Although our sample only includes sources which are radio detected, and we are unable to investigate the majority of the population which is radio undetected, our sample is dominated by AGN  which are not extremely luminous in the radio (see Figure~\ref{fig:RadOiiiLum}). Specifically, only 4.5$_{-0.5}^{+0.1}$\% and 1.0$_{-0.1}^{+0.0}$\% are above 10$^{24}$ or 10$^{25}$\,W\,Hz$^{-1}$, respectively (where the quoted range is for spectral indices varying from $\alpha=-0.8$ to $\alpha=+1.1$; \citealt{Lal10}).

\subsection{Follow-up radio observations}

\label{sec:VLAdata}
\begin{figure}
	\includegraphics[width=\columnwidth]{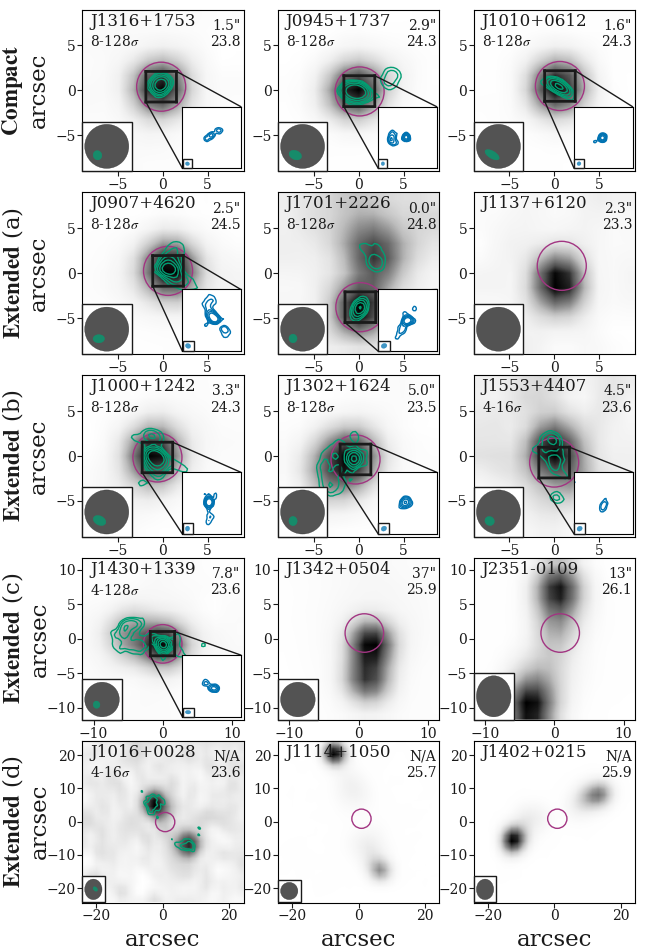}
    \caption{Examples of radio images to demonstrate our classification into compact and extended radio sources (Section~\ref{sec:radioClassification}).  Grey-scale images show 1.4\,GHz FIRST data ($\approx$5\,arcsec resolution). Where available, green contours are from our $\sim$1\,arcsec resolution 1.4\,GHz images and the insets are from our $\sim$0.3\,arcsec resolution 6\,GHz images (blue contours; Section~\ref{sec:VLAdata}). Synthesised beam(s) are represented by appropriately coloured ellipses. Labels provide each galaxy's name (\citealt{Jarvis19}, Jarvis et~al. in prep), minimum and maximum contour levels, radio sizes from FIRST ($R_{\rm Maj}$) and radio luminosities (in $\log$[W\,Hz$^{-1}]$). Magenta circles represent the SDSS fibre size. Compact radio sources (top row) are defined to have their radio emission dominated within the SDSS fibre, whilst extended sources (bottom four rows) show significant emission outside of the fibre extent.}
    \label{fig:MirandaPlot}
\end{figure}

To aid the interpretation of our results on this large sample of 2922 AGN, we also make use of follow-up radio observations that we have carried out with the Karl G. Jansky Very Large Array (VLA) that reach a higher spatial resolution than achieved by the FIRST and NVSS surveys. Example data are compared to the FIRST images in Figure~\ref{fig:MirandaPlot}. For our VLA observations we targeted a subset of the $z<0.2$ AGN from \citealt{Mullaney13} to observe at 1.4\,GHz (at $\approx$1\,arcsec resolution) and 6\,GHz (at $\approx$0.3\,arcsecond resolution).\footnote{The VLA programme IDs are: 13B$-$127; 16A$-$182 and 18A$-$300.} As in this work, the targets for these programmes were selected from \cite{Mullaney13} to be radio detected in FIRST and/or NVSS but with an additional focus on AGN with $L_{\rm [O III]}$$>$$10^{42}$\,erg\,s$^{-1}$ (see discussion in Section~\ref{sec:prevalenceResults}). The radio images typically have root-mean-square (RMS) noise values of 10--50\,$\mu$Jy. The first 10 targets from these follow-up programmes were pre-selected to have ionised outflows and are presented in \cite{Jarvis19}. The wider sample of 42, with no pre-selection on outflows, will be presented in Jarvis et~al. in prep. In Figure~\ref{fig:MirandaPlot}, we include examples of our radio images from the full sample, with our 1.4\,GHz images shown as green contours and our 6\,GHz images inset with blue contours. The observing strategy and imaging techniques are described in \citealt{Jarvis19}. In this work we make use of these images to qualitatively describe the radio morphologies and to also give an indication as to what structures might be prevalent across the wider sample (e.g., radio jets; Section~\ref{sec:radioClassification}; Figure~\ref{fig:MirandaPlot}).  

\section{Analyses}

In this work we are interested in: (1) searching for high-velocity ionised gas, indicative of outflows, by characterising the [O~{\sc iii}]$\lambda$5007 emission-line profiles and (2) relating the ionised gas kinematics to the spatial extent of the radio emission. In the following we describe how we characterise the [O~{\sc iii}] emission-line profiles (Section~\ref{sec:oiiiCharacterisation}) and constrain the spatial extent of the radio emission, defining each source as having either compact or extended radio emission (Section~\ref{sec:radioClassification}).

\subsection{Characterising the emission-line profiles}
\label{sec:oiiiCharacterisation}

\begin{figure}
	\includegraphics[width=0.9\columnwidth]{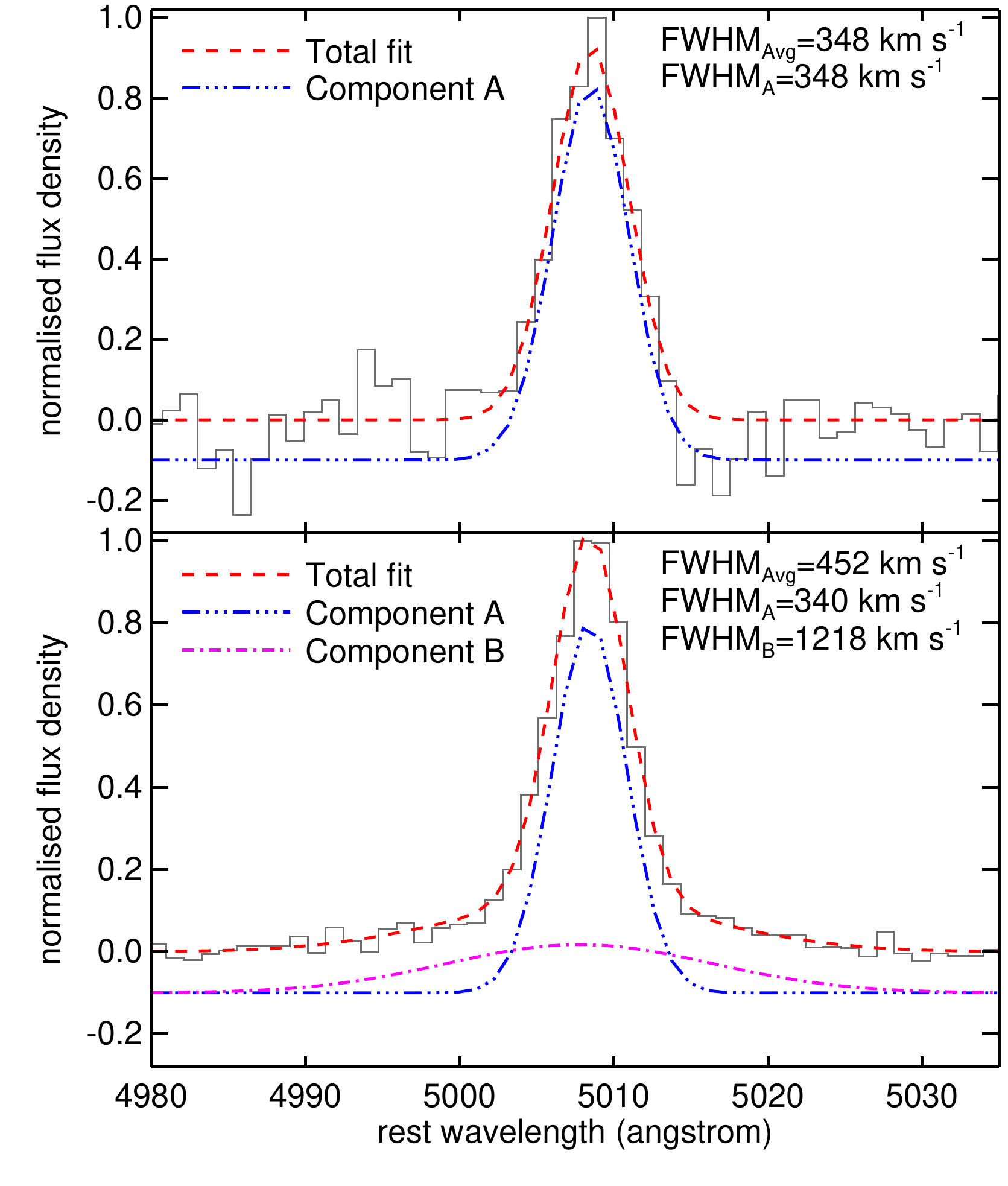}
    \caption{[O~{\sc iii}]$\lambda$5007 emission-line profiles (grey curves) and their fits (dashed curves) for an example with a single component ({\em Upper panel}) and an example with two components ({\em Lower panel}). The individual components ``A'' and ``B'' are shown as dot-dashed and three-dot-dashed curves, respectively, artificially offset by -0.1 in the y-axis. In this work we consider both the velocity widths of any identified broad components (FWHM$_{\rm B}$) and the flux-weighted average line widths of all identified components (FWHM$_{\rm Avg}$).}
    \label{fig:OIIIexamples}
\end{figure}

The velocity widths of the [O~{\sc iii}] emission-line profiles are good tracers of the ionised gas kinematics, and in particular, for identifying non-galactic motions associated with outflows \citep[e.g.,][]{Mullaney13,Liu13}. Here we make use of the two component fits to the [O~{\sc iii}] emission-line profiles provided by \cite{Mullaney13} and characterise the velocity widths in two different ways (we show two examples in Figure~\ref{fig:OIIIexamples}):
\begin{itemize}
\item The full-width-at-half-maximum of the second, broader, component (FWHM$_{\rm B}$; see bottom panel in Figure~\ref{fig:OIIIexamples}). For 2239 of the 2922 AGN in the parent sample, i.e., 77\%, such a second component is required. We discuss the prevalence of high velocity (broad) components in Section~\ref{sec:sizeTrend}.
\item  The flux-averaged FWHM of the two Gaussian components, FWHM$_{\rm Avg}$=$\sqrt{({\rm FWHM}_{A}F_{A})^2+({\rm FWHM}_{B}F_B)^2}$, where $F_{A}$ and $F_{B}$ are the fractional fluxes contained within the two fitted components, A and B. This definition has the advantage of considering lines that are fitted either with one or two Gaussians the same (e.g., sources may have broad emission-line profiles but are satisfactorily fitted with only a single Gaussian; \citealt{Mullaney13}, see upper panel in Figure~\ref{fig:OIIIexamples}).
\end{itemize}
 We take into account the corresponding uncertainties on each of these velocity width values \citep[][]{Mullaney13}, when presenting our results in Section~\ref{sec:prevalenceResults}. 

\subsection{Compact and extended radio emission}
\label{sec:radioClassification}

To separate ``compact'' from ``extended'' radio sources we determined which sources have their radio emission extent within, or outside of, the spatial extent of the SDSS fibre (i.e., 3\,arcsec diameter). This is so that we can make a connection to the observed [O~{\sc iii}] emission-line profiles seen in the fibre spectroscopy (see above). This is required for us to evaluate the radio size -- outflow connection in our sample of $\approx$3000 in the absence of spatially-resolved spectroscopy \cite[which is currently limited to considerably smaller samples; e.g.,][]{VillarMartin17,Jarvis19}. For the redshift range of our sample ($z$=0.02--0.2), 3\,arcsec corresponds to 1.2--10\,kpc; however, we note that our conclusions hold if we only consider sources $z$=0.1--0.2, for which the physical size scale varies by less than a factor of two, and if we choose a physical size cut off of 5\,kpc (Section~\ref{sec:prevalenceResults}).  

To characterise the extent of the radio emission we combine two different approaches: (1) we use radio major axis sizes ($R_{\rm maj}$) from simple elliptical Gaussian models (Section~\ref{sec:RadioSizes}) and (2) we use an automated morphological classification scheme (Section~\ref{sec:FIRSTclassifier}). It was necessary to combine both of these approaches because whilst the former method has the advantage of providing a quantitative measure of the radio sizes, and corresponding uncertainties, it has the disadvantage of missing structures that are not well characterised by a single elliptical Gaussian model. For example, there are 89 sources in our sample which are detected by NVSS but are not in the FIRST catalogue at all, largely because the emission is located in large diffuse structures or off-nuclear lobes which are missed due to the relatively small beam of FIRST (e.g., see bottom row in Figure~\ref{fig:MirandaPlot}). Furthermore, in the FIRST catalogue, the fits to the images are dominated by central cores even if there is additional extended radio structures beyond the core (e.g., see rows 2 \& 4 in Figure~\ref{fig:MirandaPlot}).

\subsubsection{Basic radio size measurements}
\label{sec:RadioSizes}

\begin{figure}
	\includegraphics[width=\columnwidth]{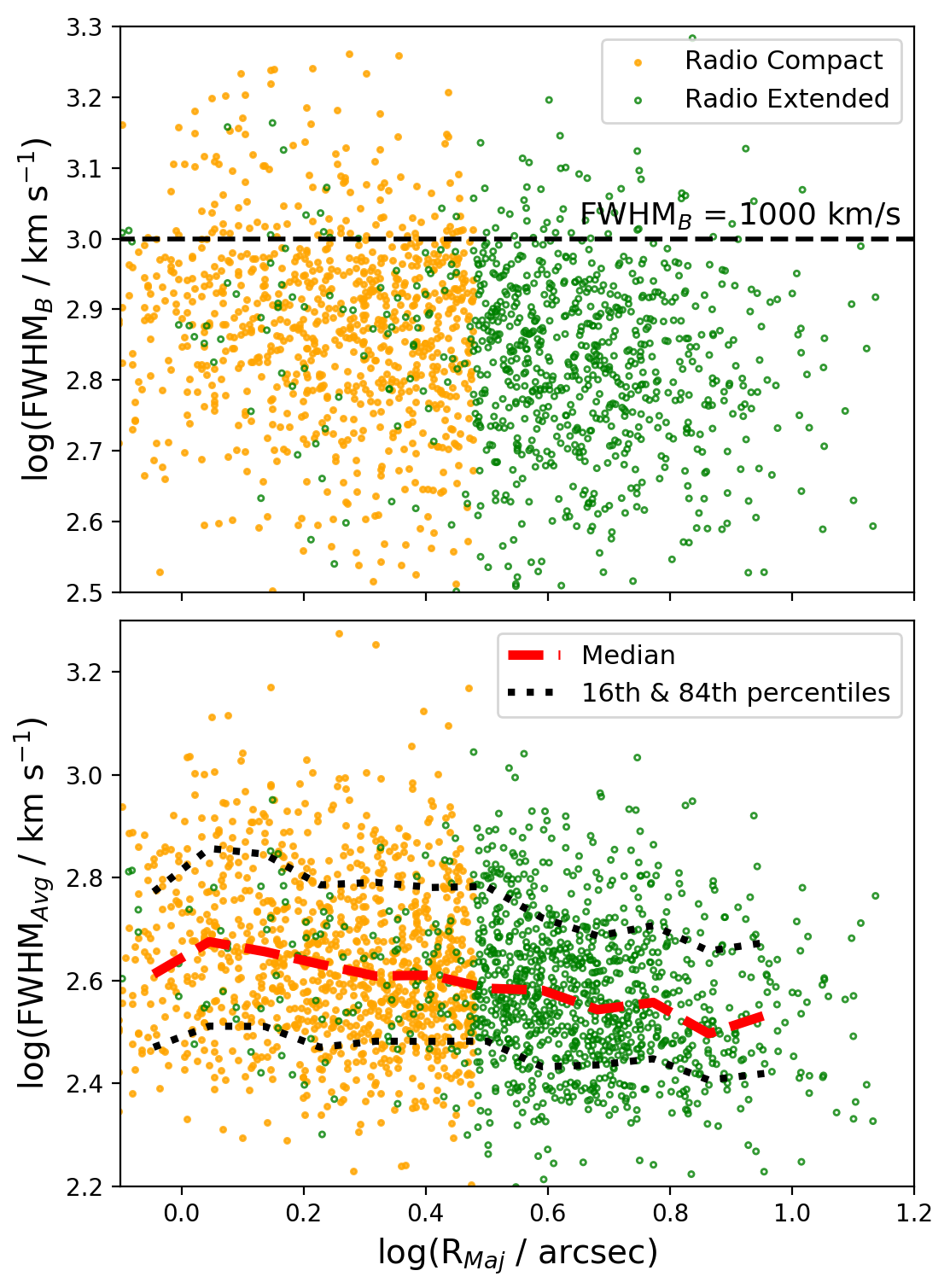}
    \caption{[O~{\sc iii}] emission-line width versus radio size, using two definitions: (1) the velocity-width of the broadest [O~{\sc iii}] emission-line component (FWHM$_{\rm B}$; top panel) and (2) the flux-weighted average velocity width of two emission-line components (bottom panel; FWHM$_{\rm Avg}$). Symbols are colour-coded as in Figure~\ref{fig:RadOiiiLum} with the addition of the curve showing the running median (in 0.1\,dex bins) and the corresponding 16th and 84th percentiles in the bottom panel. The small number of ``radio extended'' sources with apparently small sizes are due to extended radio structures beyond the core (see Section~\ref{sec:radioClassification}). We see a weak trend where the largest, $\approx$8\,arcsec sources, have a 0.1--0.2\,dex smaller velocity widths, on average, compared to the smaller $\approx$1\,arcsec sources. Compact radio sources are more likely to have the broadest emission-line components  (e.g., FWHM$_{\rm B}>$1000km\,s$^{-1}$; above dashed line; Section~\ref{sec:sizeTrend}).}  
    \label{fig:Size_vs_fwhm}
\end{figure}

\citet{Helfand15} fit an elliptical Gaussian model for each source detected in their FIRST catalogue. We make use of these major axis sizes ($R_{\rm Maj}$) which have been deconvolved for the point-spread function (or `beam'), noting that 89 sources (3\% of the sample) do not have any constraints on $R_{\rm Maj}$ because they only have radio detections in the NVSS catalogue (see above).\footnote{Noise can cause the $R_{\rm Maj}$ sizes (before deconvolution) to be smaller than the beam (see \citealt{Helfand15}). For the 618 cases in our sample the corresponding deconvolved sizes are assumed to be zero (example in second row of Figure~\ref{fig:MirandaPlot}).} The uncertainties in $R_{\rm Maj}$ depend on the signal-to-noise ratio (SNR), following approximately $\sigma$($R_{\rm Maj}$)= 10$\times$(1/SNR + 1/75)\,arcsec, where SNR=($F_{\rm peak}$-0.25)/RMS, $F_{\rm peak}$ is the peak flux density and RMS is the root-mean-square noise of the image (\citealt{Helfand15}). The signal-to-noise ratios are not simply $F_{\rm peak}$/RMS, because of the applied {\sc clean} bias correction to the peak flux density. Importantly, as demonstrated by this equation, it is possible to obtain sizes with reasonable uncertainties well below the size of the nominal point-spread-function when the sources are detected with high signal-to-noise ratios. For example, a source with a SNR$=$10 has a size uncertainty of $\approx$1\,arcsec. We take into account the size uncertainties when presenting our results in Section~\ref{sec:prevalenceResults}. We plot the $R_{\rm Maj}$ values in Figure~\ref{fig:Size_vs_fwhm}.


We can also define how extended a source is in the FIRST images based on the ratio of peak and integrated flux densities following $\theta=\sqrt{F_{\rm Int}/F_{\rm Peak}}$ \citep{Kimball08}, where $F_{\rm Int}$ is the integrated flux from the elliptical Gaussian models (\citealt{Helfand15}). These $\theta$ values describe how resolved a source is because a larger $F_{\rm Int}$/$F_{\rm Peak}$ ratio implies that the extended radio emission (e.g., radio lobes) contributes more to the total radio flux. A source in FIRST can typically be considered `extended' if $\theta\ge1.06$ \citep[][]{Kimball08}. We find that $R_{\rm Maj}$ and $\theta$ are well correlated for our sample, with a correlation coefficient value of 0.91, and the overall conclusions presented throughout this work are insensitive to the choice of setting $R_{\rm Maj}>3$\,arcsec versus $\theta>1.06$ to define a source as ``radio extended''.

\subsubsection{Morphological classification}
\label{sec:FIRSTclassifier}

As previously described, for a complete characterisation of which sources are compact or extended in the radio, it is important to apply a morphological classification in addition to simple sizes (e.g., to identify core-lobe structures). To do this we make use of the ``FIRST Classifier'', presented in \citet{Alhassan18}, which is an automatic morphological classification tool applied to FIRST images that uses a trained, deep Convolutional Neural Network model. The code was ``trained'' using a set of radio sources with known classifications. Sources are classified as FRI, FRII (Fanaroff and Riley class I and II; \citealt{Fanaroff74}), Bent (determined to have a more complex, ``bent" nature) or Compact. For the current study we are not interested in the individual classifications of FRI, FRII or Bent, just if the radio emission is extended or compact. The model achieves an overall accuracy of 97\,per\,cent based on control samples \citep{Alhassan18}. 

 We performed our own random visual inspection to verify that the morphological classifications returned by the FIRST Classifier were reliable. Furthermore, we could make use of the subset of our sample which are covered by our follow-up higher resolution radio observations (Section~\ref{sec:VLAdata}; Figure~\ref{fig:MirandaPlot}). Indeed, the FIRST Classifier is successful at identifying extended radio structures (e.g., those sources in the second and fourth row of Figure~\ref{fig:MirandaPlot}). Nonetheless, sources which are smoothly extended beyond 3$^{\prime\prime}$ in FIRST, without clear distinct morphological structures can be miss-classified as compact by the FIRST Classifier (e.g., see third row of Figure~\ref{fig:MirandaPlot}). Therefore, we found that a combination of using $R_{\rm Maj}$ and the results of the FIRST Classifier was required to robustly classify all of the sources in our sample as either ``radio compact" or ``radio extended".

 \subsubsection{Final classification into compact and extended radio sources}

 As described above, we wish to define compact sources as those where the radio emission is concentrated {\em within} the SDSS fibre (i.e., $\lesssim$3\,arcsec), and define sources as extended otherwise. Based on the measurements described in the previous two sub-sections, we apply the following criteria to separate the two populations:
\begin{itemize}
    \item {\bf Compact:} $R_{\rm Maj}\le3$\,arcsec and not identified as extended by the FIRST Classifier (1620 sources; i.e. see first row in Figure~\ref{fig:MirandaPlot}).
    \item {\bf Extended (a)}: Sources which have $R_{\rm Maj}\le3$\,arcsec but identified as extended by the FIRST Classifier (246\,sources). Visual inspection shows that these targets typically have strong cores (which results in the small sizes from simple 2D-Gaussian fits) but with additional extended structures expanding beyond the core which are picked up the the FIRST Classifier (see second row in Figure~\ref{fig:MirandaPlot}). 
    \item {\bf Extended (b):}  $R_{\rm Maj}>3$\,arcsec but not classified as extended by the FIRST classifier (679 sources). Visual inspection reveals that these targets are large but do not have clear discernible features which would be picked up by the FIRST classifier (see third row in Figure~\ref{fig:MirandaPlot}). 
    \item {\bf Extended (c):} $R_{\rm Maj}>3$\,arcsec and classified as extended by the FIRST Classifier. These sources are clearly extended with morphological structures on large scales (288\,sources; see fourth row in Figure~\ref{fig:MirandaPlot}).
    \item {\bf Extended (d):} Those not detected in the FIRST catalogue but are identified in NVSS (89 sources). Visual inspection verifies that these targets are large radio sources, which are usually dominated by symmetrical lobes (see bottom row of Figure~\ref{fig:MirandaPlot}).
\end{itemize}

Using these criteria 1302 sources are classified as extended and 1620 are classified as compact in the radio. A visual inspection of the FIRST images, in combination with our follow-up radio observations (Figure~\ref{fig:MirandaPlot}) provides verification that these criteria are effective at separating sources where the radio emission is dominated within 3\,arcsec or extends beyond. Unsurprisingly there are a few ambiguous cases within the full sample of 2922. For example, from our follow-up radio observations, J0945$+$1737 in the top row of Figure~\ref{fig:MirandaPlot} is known to have a weak radio structure beyond the central 3\,arcsec; however, this constitutes only $\approx$9\% of the total radio emission and we see that the majority of the radio emission is due to a compact $\lesssim$2\,kpc central core and radio jet (\citealt{Jarvis19}). Overall we feel confident that for this statistical study our classification into compact and extended radio sources is sufficient and is likely the best that can be achieved with the current radio surveys.

The radio and [O~{\sc iii}] luminosities for the compact and extended samples are represented as solid and hollow symbols, respectively, in Figure~\ref{fig:RadOiiiLum}. Their median radio luminosities are $\log[L_{\rm 1.4GHz}$/W\,Hz$^{-1}$]=22.80 and 22.67, respectively, with a standard deviation 0.7\,dex in both cases. The median [O~{\sc iii}] luminosities are $\log[L_{\rm [O III]}$/erg\,s$^{-1}$]=41.18 and 41.03, respectively, both with a standard deviation of 0.7\,dex. Although there are only small, $\approx$40\%, differences in the median radio and [O~{\sc iii}] luminosities of the two populations, we note that we repeat all of our experiments using individual 1\,dex bins of radio and [O~{\sc iii}] luminosity to account for these differences, in Section~\ref{sec:prevalenceResults}. 

With the final classifications of compact versus extended radio emission we can now explore the relationship between radio size and ionised gas kinematics.



 \section{Results and Discussion}
 
\begin{center}
\begin{table*}
\resizebox{0.82\paperwidth}{!}{
\begin{tabular}{cccccccccccc}
\hline
$z$ &  $L_{\rm 1.4 GHz}$    & $L_{\rm [O III]}$ & N$_{\rm Comp.}$ & N$_{\rm Ext.}$ &  p-value & p-value& Compact  & Extended & Compact more\\ 
 &  ($\log$[W/Hz])   & ($\log$[erg/s]) & & &  FWHM$_{\rm Avg}$ & FWHM$_{\rm B}$ & \%$>$1000\,km/s & \%$>$1000\,km/s & extreme outflows?\\ 
(1) &  (2)   & (3) & (4) & (5) & (6) & (7) & (8) & (9) & (10)  \\ 
\hline\hline
0.02--0.2 & All & All & 1620 & 1302 & 3.9$\times$10$^{-14}$($<$6.6$\times$10$^{-9}$) & 2.7$\times$10$^{-12}$ ($<$1.7$\times$10$^{-6}$) & 11 ($<$13)  & 5.7 ($<$8.5) & Y\\
0.1--0.2 & All & All & 813 & 584 & 2.6$\times$10$^{-8}$ ($<$4.1$\times$10$^{-4}$) & 6.4$\times$10$^{-5}$ ($<$1.6$\times$10$^{-2}$) & 15 ($<$19) & 9.1 ($<$13) & Y\\
0.02--0.2 & 21.5--22.5 & All & 433 & 460 & 2.2$\times$10$^{-2}$ ($<$4.1$\times$10$^{-1}$) & 3.2$\times$10$^{-3}$ ($<$4.3$\times$10$^{-1}$) & 4.0 ($<$7.4) & 2.2 ($<$5.9) & ?\\
0.02--0.2 & 22.5--23.5 & All & 958 & 723 & 8.4$\times$10$^{-6}$ ($<$3.5$\times$10$^{-3}$) & 3.8$\times$10$^{-6}$ ($<$9.1$\times$10$^{-3}$) & 11 ($<$14) & 7.9 ($<$11) & Y\\
0.02--0.2 & 23.5--24.5 & All & 180 & 60 & 4.4$\times$10$^{-4}$ ($<$3.0$\times$10$^{-2}$) & 4.4$\times$10$^{-3}$ ($<$5.9$\times$10$^{-2}$) & 25 ($<$28) & 6.7 ($<$14)& Y\\
0.02--0.2 & All & 39.5--40.5 & 274 & 258 &  6.1$\times$10$^{-4}$ ($<$2.0$\times$10$^{-1}$) & 3.4$\times$10$^{-1}$ ($<$1.0$\times$10$^{-0}$) & 1.1 ($<$4.9) & 0.0 ($<$5.4) & ?\\
0.02--0.2 & All & 40.5--41.5 & 869 & 741 & 4.4$\times$10$^{-5}$ ($<$3.0$\times$10$^{-2}$) & 9.6$\times$10$^{-6}$ ($<$3.8$\times$10$^{-2}$) & 7.8 ($<$11) & 5.1 ($<$8.6) & ?\\
0.02--0.2 & All & 41.5--42.5 & 440 & 264 &  5.2$\times$10$^{-6}$ ($<$7.4$\times$10$^{-4}$) & 6.1$\times$10$^{-5}$ ($<$1.3$\times$10$^{-2}$) & 22 ($<$25) & 13 ($<$16) & Y\\
\hline
\end{tabular}}
\caption{Results of comparing the [O~{\sc iii}] emission-line profiles for AGN with compact versus extended radio emission for various subsets of the sample. Column definitions are as follows: (1)--(3) ranges of redshift, radio luminosity and [O~{\sc iii}] luminosity used in each subset; (4) \& (5) number of AGN in the subset with compact and extended radio emission, respectively; (6) \& (7) p-values from two-sided KS-tests for comparing the distributions of FWHM$_{Avg}$ and FWHM$_{B}$ (see Section~\ref{sec:prevalenceResults}), respectively, for the compact versus extended radio sources (p-values $<$5$\times$10$^{-2}$ means statistically different distributions); (8) \& (9) the percentage of compact and extended radio sources, respectively, with a broad [O~{\sc iii}] emission-line component FWHM$_{\rm B}>$1000\,km\,s$^{-1}$ (see Section~\ref{sec:sizeTrend}) (for columns (6)--(9) in brackets, with give the 99.7th percentile of the corresponding values from a 10$^{4}$ run Monte Carlo simulation where we randomly perturbed all relevant values by their uncertainties; see Section~\ref{sec:prevalenceResults}); (10) based on the 99.7th percentiles: `Y' if it is confident that compact radio sources have a higher prevalence of FWHM$_{B}$ $>$1000\,km\,s$^{-1}$ components, otherwise '?' is shown.}
\label{Tab:results}
\end{table*}
\end{center}


To investigate the relationship between the prevalence of extreme ionised outflows and the size of the radio emission in AGN host galaxies, we have constructed a sample of 2922 $z$=0.02--0.2, spectroscopically-identified AGN which are detected in 1.4\,GHz radio images (Section~\ref{sec:sample}; Figure~\ref{fig:RadOiiiLum}). Using a combination of direct size measurements and morphological classifications we have identified the sources which are ``compact'' versus ``extended'' in the radio, based upon if the radio emission is dominated within or outside of $\approx3$\,arcsec (or $\approx$5\,kpc at the average redshift; Section~\ref{sec:radioClassification}; Figure~\ref{fig:MirandaPlot}). Two Gaussian component fits to the [O~{\sc iii}] emission-line profiles have been used to characterise the ionised gas velocities using: (1) the width of any identified broad emission-line components (FWHM$_{\rm B}$) and (2) the flux-averaged width of the two components (FWHM$_{\rm Avg}$; Section~\ref{sec:oiiiCharacterisation}; Figure~\ref{fig:OIIIexamples}). Here we present our results on the trend between radio size and ionised gas velocities (Section~\ref{sec:sizeTrend}) and with the prevalence of ionised outflows (Section~\ref{sec:prevalenceResults}), before discussing the implication of our results in the context of AGN feedback and how the outflows are driven (Section~\ref{sec:implications}). The quantitative results of our analyses are presented in Table~\ref{Tab:results}.

 \subsection{Trends between ionised gas velocities and radio sizes}
 \label{sec:sizeTrend}
 
In Figure~\ref{fig:Size_vs_fwhm} we plot the FWHM of the [O~{\sc iii}] emission lines versus radio size ($R_{\rm Maj}$) for the AGN in our sample. In the bottom panel, we show the same but using FWHM$_{\rm Avg}$, which has the advantage of being defined for those targets with one {\rm or} two Gaussian component fits to the [O~{\sc iii}] profiles (see Section~\ref{sec:oiiiCharacterisation}) and gives a sense of the flux-weighted average ionised gas velocities inside the galaxy (covered by the spectroscopic fibres). It can be seen that there is a general trend that the largest radio sources typically have lower ionised gas velocities: the average FWHM$_{\rm Avg}$ drops by 35\% from $R_{\rm Maj}$=1\,arcsec to $R_{\rm Maj}$=8\,arcsec. More easily interpreted is the top panel of Figure~\ref{fig:Size_vs_fwhm}, where the percentage of targets which have a very high velocity component of FWHM$_B$>1000\,km\,s$^{-1}$, indicative of extreme outflows, is higher for the smaller radio sources (see dashed line). Specifically, for targets with $R_{\rm Maj}<$3\,arcsec, 10.7\% exhibit [O~{\sc iii}] emission-line components with FWHM$_B$>1000\,km\,s$^{-1}$, whilst such features are half as common (5.3\%) for targets with $R_{\rm Maj}>$3\,arcsec. 
 
 
 One important limitation of using the radio size measurements, $R_{\rm Maj}$, in Figure~\ref{fig:Size_vs_fwhm} is that it fails to properly capture sources with large extended radio structures, which is why sources classified as ``extended'' (green data points) can have apparently small sizes (see Section~\ref{sec:radioClassification}). Furthermore, until this point we have not considered the uncertainties on the size or velocity measurements. Therefore, in the following subsection we quantify the relationship between radio size and the prevalence of extreme outflows further, using our careful classifications of compact versus extended radio emission (Section~\ref{sec:radioClassification}) and accounting for the uncertainties. 
 
\subsection{Extreme outflows are more prevalent with compact radio emission}
\label{sec:prevalenceResults}

\begin{figure}
    \includegraphics[width=\columnwidth]{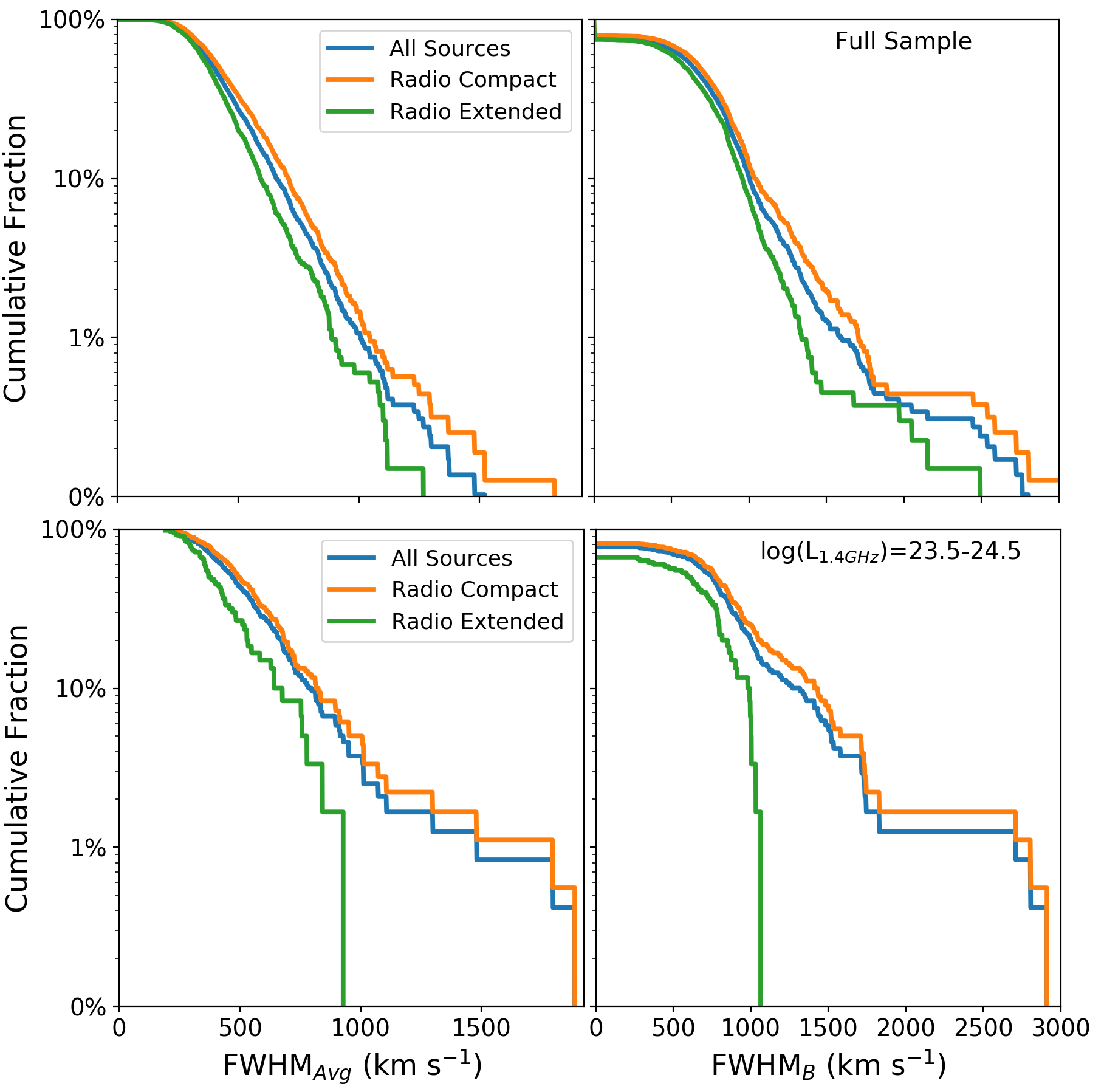}
    \caption{Fraction of AGN with [O~{\sc iii}] FWHM greater than a given value, for both FWHM$_{\rm Avg}$ (left two panels) and FWHM$_{\rm B}$ (right two panels). The top two panels show the full sample of this work and the bottom two panels show the subset with radio luminosities of $\log[L_{\rm 1.4GHz}$/W Hz$^{-1}]$=23.5--24.5. The cumulative distributions are shown for all the sources in each bin (blue curves) and split by compact and extended radio sources (orange and green curves, respectively). Extreme ionised gas velocities are more prevalent when the radio emission is compact; for example, components with FWHM$_{\rm B}>$1000\,km\,s$^{-1}$ are four times as likely in compact sources when considering the $\log[L_{\rm 1.4GHz}$/W Hz$^{-1}]$=23.5--24.5 sample (bottom right panel).}
    \label{fig:CumulFrac}
\end{figure}

In the top two panels of Figure~\ref{fig:CumulFrac} we show the cumulative distribution of FWHM$_{\rm Avg}$ and FWHM$_{\rm B}$ for the full AGN sample of this work. Sources which are radio compact (orange curves) typically have higher ionised gas velocities than sources which are radio extended (green curves). A two-sided KS-test (Kolmogorov–Smirnov test) comparing the radio compact sources to the radio extended sources reveals that they have FWHM$_{\rm Avg}$ and FWHM$_{\rm B}$ distributions which are {\em not} consistent with each other, with p-values of 3.9$\times$10$^{-14}$ and 2.7$\times$10$^{-12}$, respectively.\footnote{Where a p-value<0.05 rejects the null hypothesis that the two samples are drawn from the same distribution.} Furthermore, we found that compact sources are twice as likely to have FWHM$_{\rm B}$>1000\,km\,s$^{-1}$ emission-line components compared to extended radio sources, with 11\% versus 5.7\% exhibiting such components, respectively. Such differences between compact and extended radio sources also remain if we focus on a narrower redshift range of $z$=0.1--0.2 so that there is only a factor of two variation in the physical size scale covered by the SDSS fibre.\footnote{We note that  if we use 5\,kpc, as opposed to 3\,arcsec, to define ``radio extended'' in the $z$=0.1--0.2 bin we obtain extreme outflow detection fractions of 15.7\% and 10.5\% for compact and extended radio sources, respectively.} For a full breakdown of these results see the top two rows in Table~\ref{Tab:results}. 

As mentioned previously it is important for us to take into account the uncertainties on both the radio size measurements and emission-line velocity width measurements. Therefore, we used a Monte Carlo approach to generate 10$^{4}$ sets of $R_{\rm Maj}$, FWHM$_{B}$ and FWHM$_{\rm Avg}$ values for the 2922 targets in our sample. To do this, we randomly perturbed the true values by their uncertainties using a normal distribution with a width equal to the measurement errors. For each of the 10$^{4}$ sets, we re-classified the full sample as ``compact'' or ``extended'', following Section~\ref{sec:radioClassification}, and re-performed the two-sided KS-tests on the  FWHM$_{B}$ and FWHM$_{\rm Avg}$ distributions. From this exercise we found that the 99.73 percentiles of the p-values from the 10$^{4}$ Monte Carlo runs are 6.6$\times$10$^{-9}$ and 1.7$\times$10$^{-6}$ for FWHM$_{B}$ and FWHM$_{\rm Avg}$, respectively. This shows that when folding in the uncertainties the $\approx$3$\sigma$ maximum p-values still reveal that the ``compact'' and ``extended'' radio sources have FWHM$_{\rm Avg}$ and FWHM$_{\rm B}$ distributions which are {\em not} consistent with each other. Furthermore, using these same Monte Carlo runs, at most, 8.5\% of the compact radio sources have FWHM$_{B}$>1000\,km\,s$^{-1}$ (again using the 99.73 percentile), which is still a smaller fraction than the 11\% observed in the compact radio sources (see Table~\ref{Tab:results}). 

Another important consideration when interpreting our results is to confirm that the radio size/morphology is the driving physical parameter on the different prevalence of extreme ionised outflows and that it is not driven by the underlying [O~{\sc iii}] or radio luminosities \citep[Figure~\ref{fig:RadOiiiLum},][]{Mullaney13}. To test for this we repeat the above calculations and Monte Carlo simulations but we split the full sample into 1\,dex bins of [O~{\sc iii}] luminosity and radio luminosity. We only considered bins with $>$50 sources, which meant we could explore the luminosity ranges of $\log[L_{\rm 1.4GHz}$/W\,Hz$^{-1}$]=21.5--24.5 and $\log[L_{\rm [O III]}$/erg\,s$^{-1}$]=39.5--42.5, split into three bins in both cases. The full results of these tests can be found in Table~\ref{Tab:results}. We find that for all radio luminosity bins, except the lowest bin, we can be confident that the prevalence of extreme outflows (i.e., components with FWHM$_{B}$>1000\,km\,s$^{-1}$) is higher for the compact sources (even after accounting for the uncertainties; see Table~\ref{Tab:results}). Furthermore, in nearly all luminosity bins the p-values consistently show that the compact and extended radio sources do not have the same distributions of FWHM$_{\rm Avg}$ or FWHM$_{\rm B}$, even when considering the uncertainties during the Monte Carlo runs. The exceptions to this are the two lowest luminosity bins of $\log[L_{\rm 1.4GHz}$/W\,Hz$^{-1}$]=21.5--22.5 and $\log[L_{\rm [O III]}$/erg\,s$^{-1}$]=39.5--40.5, where the AGN are likely to be particularly weak and/or star-formation processes dominate the observed radio luminosities.

Exploring the effect of luminosity further we find that the difference in ionised gas velocities, between compact and extended radio sources, is most significant in our largest radio luminosity range considered of $\log[L_{\rm 1.4GHz}$/W\,Hz$^{-1}$]=23.5--24.5. The FWHM cumulative distributions for this sub-sample are shown in the bottom two panels of Figure~\ref{fig:CumulFrac}. This result is quantified by the fact that the prevalence of FWHM$_{B}$>1000\,km\,s$^{-1}$ components is almost {\em four times} higher in the compact versus extended radio sources, compared to a factor of two for the full population. In this radio luminosity range, AGN are generally accepted to dominate the radio emission at low redshift \citep[e.g.,][]{Kimball08,Condon13,Mancuso17}. Following \citealt{Kennicutt12}, if we assumed the radio luminosities were all from star formation for the range $\log[L_{\rm 1.4GHz}$/W\,Hz$^{-1}$]=23.5--24.5, this would correspond to star-formation rates of $\approx$200--2000\,M$_{\odot}$\,yr$^{-1}$. At these redshifts, it would be extremely unlikely for more than one or two of the sources to have such high star-formation rates; for example, X-ray AGN have an average SFR of $\approx$1--8\,M$_{\odot}$\,yr$^{-1}$ at $z\lesssim$0.5 \citep[][]{Stanley15,Shimizu17}. Even more importantly, follow-up observations of subsets of the sample in this luminosity range show that star-formation is very unlikely to dominate in most cases due to the observed collimated radio structures and very high radio excess (Figure~\ref{fig:MirandaPlot}; \citealt{Jarvis19}). However, since these high-resolution radio observations (Section~\ref{sec:VLAdata}) only represent the [O~{\sc iii}] and radio luminosity bins where the difference between compact and extended radio sources is strongest, further high-resolution radio observations are required to establish the origin of radio emission in the overall population \citep[also see][]{Panessa19}. 

Finally, we also considered the possibility that Type~1 AGN may bias the results due to projection effects that could potentially lead to Doppler boosted radio emission, seemingly more compact radio emission and/or higher observed outflow velocities. Therefore, we repeated our analyses only including the Type~2 (Type~1) AGN and found that 7.5\% (27.4\%) of the radio compact sources show a $>$1000 km/s broad [O~{\sc iii}] component and 3.6\% (15.3\%) of the extended sources do. The Type~2 only sample has a lower detection fraction of extreme outflows, consistent with the results presented in \cite{Mullaney13} that Type~1 AGN typically have higher observed [O~{\sc iii}] velocities, likely due to projection effects. However, this test shows that Type~2 AGN alone show a consistent result with the overall sample that extreme outflows are roughly twice as common when the radio emission is compact. This is still true if we focus only on the Type~2 AGN in our highest radio luminosity bin, where our result is strongest (see Figure~\ref{fig:CumulFrac}), finding that 21.1\% for the compact radio sources, compared to 8.5\% for the extended radio sources exhibit extreme [O~{\sc iii}] components of $>$1000\,km\,s$^{-1}$. It will be interesting to explore, and to further understand, differences between Type~1 AGN and Type~2 AGN in the future using larger samples which are complete down to lower radio luminosities. 

Overall, we conclude that the prevalence of extreme ionised outflows is highest when the radio emission is compact and the AGN are clearly the dominating source of radio emission (i.e., $\log[L_{\rm 1.4GHz}$/W\,Hz$^{-1}$]=23.5--24.5). Since we only have small numbers of sources at higher radio luminosities, we are unable to extend our results to $\log[L_{\rm 1.4GHz}$/W\,Hz$^{-1}$]$>$24.5.

\subsection{The implications of our results}
\label{sec:implications}
We find that extreme ionised outflows observed in SDSS spectroscopy for $z<0.2$ AGN are more common when the radio emission is concentrated within the spatial extent of the spectroscopic fibre. This re-enforces the idea that there is a connection between radio emission and outflows in AGN \citep[e.g.,][]{Mullaney13, Zakamska14}. For the sample at hand (i.e., where the radio emission is detected; Section~\ref{sec:sample}), we suggest that this is not driven by star-formation processes because the result is strongest when AGN will be dominating the radio emission. Our result shows that the connection between outflows and compact radio emission for extremely radio bright AGN, found by \cite{Holt08}, is also found in AGN which are not extremely radio luminous and, hence, are more representative of the overall population.

If the radio emission is tracing the extent of jets, our result could imply that we can see the effect of jet-ISM interactions from young radio sources or low-power frustrated jets that will never escape the host galaxy \citep[e.g.,][]{vanBreugel84,ODea91,ODea98,Morganti17,Bicknell18}. Where larger scale jets are depositing their energy outside of the region covered by the spectroscopic fibre, the spectroscopic measurements do not cover the physical region of jet-ISM interactions. 

In favour of the jet scenario, we see collimated jet-like features (including hot spots and `bent' jets) in our follow-up high resolution radio observations (Figure~\ref{fig:MirandaPlot}; \citealt{Jarvis19}). Furthermore, spatially-resolved spectroscopic observations reveal jet-ISM interactions in sources with a range of jet powers, particularly on the scale of the galaxy bulges \citep[i.e., a few kiloparsec; ][]{Tadhunter16,VillarMartin17,Jarvis19,Husemann19a,Husemann19b}. Indeed, spatially-resolved spectroscopic observations are also crucial for de-coupling less extreme ionised outflows from galactic kinematics; here we are only able to confidently identify the most extreme cases.

Cutting-edge simulations show that compact jets interacting with a clumpy ISM may be a crucial aspect of `AGN feedback' and possibly the most efficient mechanism for driving powerful outflows \citep[e.g.,][]{Wagner12,Mukherjee16,Bicknell18,Cielo+18}. Alternatively, the increased prevalence of extreme outflows for compact radio emission may be because quasar-driven winds drive the ionised outflows and simultaneously shock the ISM to produce radio emission in the same region of the galaxy \citep[][]{Wagner13,Zakamska14,Nims15,Zakamska16,Wagner16,Hwang18}. This scenario could become indistinguishable from those driven by jets, especially in the cases where the jets become disrupted and are more diffuse \citep[][]{Wagner13,Alexandroff+16}. More theoretical and observational work on larger samples is required to distinguish between these two scenarios. 
 
\section{Conclusions}
We have used a sample of 2922 $z$$=$$0.02$--0.2 AGN, spectroscopically identified in SDSS, with a radio detection in FIRST and/or NVSS, to investigate the relationship between ionised outflows and the spatial extent of the radio emission. We made use of two component fits to the [O~{\sc iii}]$\lambda$5007 emission-line profiles to characterise the velocity widths, considering both the width of any identified broad components (FWHM$_{\rm B}$) and the flux-weighted average width of the two components (FWHM$_{\rm Avg}$; see Figure~\ref{fig:OIIIexamples}). To characterise the radio sizes we considered both major-axis sizes from two-dimensional Gaussian fits (deconvolved for the beam) and an automated morphological classification routine (see Figure~\ref{fig:MirandaPlot}). We find that:
\begin{itemize}
    \item Except for the AGN with the lowest [O~{\sc iii}] and radio luminosities (i.e., $\log[L_{\rm [O III]}$/erg\,s$^{-1}$]$<$40.5; $\log[L_{\rm 1.4 GHz}$/W\,Hz$^{-1}$]$<$22.5) the radio compact and radio extended sources have statistically different distributions of ionised gas kinematics. Compact radio sources tend to have broader emission-line profiles on average (Section~\ref{sec:sizeTrend}; Figure~\ref{fig:Size_vs_fwhm}).
    \item When the radio emission is confined within 3$^{\prime\prime}$ (i.e., within the SDSS fibre), equivalent to $\lesssim$5\,kpc at the median redshift, broad [O~{\sc iii}] emission-line components with FWHM$_{\rm B}$$>$$1000$\,km\,s$^{-1}$, indicative of high-velocity outflows, are twice as prevalent (Figure~\ref{fig:CumulFrac}). 
    \item Extreme outflow components (FWHM$_{\rm B}$$>$$1000$\,km\,s$^{-1}$) are four-times more prevalent when only considering the sources with moderate radio luminosities (i.e., $\log[L_{\rm 1.4 GHz}$/W\,Hz$^{-1}$]=23.5--24.5), where AGN are most likely to be the dominant source of radio emission. Follow-up sub-kpc resolution radio observations of a subset of the sample, in this luminosity range, reveal a high prevalence of moderate power jets and lobes (\citealt{Jarvis19}; Figure~\ref{fig:MirandaPlot}). We are too limited in source statistics to make strong conclusions about higher radio-luminosity AGN. 
\end{itemize}

Our results add to the growing body of evidence that there is a strong connection between the presence of ionised outflows and the radio emission in AGN host galaxies (e.g., \citealt{Mullaney13,VillarMartin14,Holt08,Zakamska14,Hwang18,Jarvis19}). We find that extreme ionised outflows are more prevalent when the radio emission is compact even for AGN which are not extremely radio luminous, as had previously been seen in the most powerful ``radio loud'' AGN \citep[][]{Holt08}. Follow-up high resolution observations of subsets of targets imply that compact, low-power radio jets, young or frustrated by interactions with the host galaxy ISM (Figure~\ref{fig:MirandaPlot}; \citealt{Jarvis19}) may be responsible for the high-velocity ionised gas, inline with some recent model predictions (e.g., \citealt{Mukherjee18}). However, we can not rule out other possible processes, such as nuclear wide-angle winds, that contribute to producing the radio emission and outflows in the wider sample (e.g., see \citealt{Zakamska16}). High-resolution observations of larger samples will help determine the relative contribution of these different processes.

This work was limited to AGN with radio detections in NVSS and/or FIRST which are relatively shallow. On-going and future, deep and large-area multi-frequency radio surveys such as: VLASS (\citealt{Lacy19}); those with LOFAR (\citealt{Smith16,Shimwell17}) and eventually those with the SKA, that are combined with spectroscopic information will be crucial to unravelling a complete picture of the origin of radio emission in AGN and to further establish the physical processes behind AGN--host galaxy interactions.

\section*{Acknowledgements}
S.M. acknowledges support from an internship provided by the European Southern Observatory's Office for Science. We thank the referee, Prof. Villar-Martin, for constructive comments which helped us to improve the clarity of the manuscript.

\bibliographystyle{aa} 

\end{document}